\documentclass[12pt] {article}
\pdfoutput=1
\usepackage[hyphens,spaces,obeyspaces]{url}
\usepackage{setspace}

\usepackage{epsfig}
\usepackage{amssymb}
\usepackage{amsmath}
\usepackage{multirow}
\usepackage{setspace}
\usepackage{datetime}
\usepackage{amsmath}
\usepackage{algorithm}
\usepackage{algpseudocode}
\usepackage{graphicx}
\usepackage{lipsum}
\usepackage{acronym}
\usepackage{array}
\usepackage{graphicx}
\usepackage{cite}
\usepackage{graphicx}
\usepackage{epstopdf}
\usepackage{caption}
\usepackage{balance}
\usepackage{fancybox}
\usepackage{colortbl}
\usepackage{array}
\usepackage{mystyle}
\usepackage{multirow}
\usepackage[obeyspaces,hyphens,spaces]{url}
\usepackage{threeparttable}
\usepackage[english]{babel}
\usepackage{longtable}
\usepackage{listings, lstautogobble}
\usepackage{booktabs}
\usepackage{threeparttable}
\usepackage[table,xcdraw]{xcolor}
\usepackage{wrapfig}
\usepackage{subfigure}
\usepackage{enumitem}
\usepackage{paralist}

\usepackage{amsthm}
\AtBeginEnvironment{thm}{\footnotesize}
\AtBeginEnvironment{defn}{\footnotesize}
\providecommand{\keywords}[1]{\textbf{\textit{Keywords---}} #1}

\usepackage{stackengine}
\newsavebox\mybox

\usepackage{hyperref}
\usepackage[framemethod=tikz]{mdframed}
\usepackage{enumitem}
\theoremstyle{definition}
\newtheorem{thm}{\small{Theorem}}
\newtheorem{defn}{Definition}
\AtBeginEnvironment{thm}{\footnotesize}
\AtBeginEnvironment{defn}{\footnotesize}

{\begin{list}{}%
         {\setlength{\leftmargin}{#1}}%
         \item[]%
}
{\end{list}}

\newdateformat{monthyeardate}{%
  \monthname[\THEMONTH], \THEYEAR}
\begin{document}
\sloppy

\title{\Large{\textbf{Integrating DFT and DRBD Formalizations in HOL4} }}
\author{
Yassmeen~Elderhalli, Osman~Hasan, and Sofi\`ene~Tahar\vspace*{2em}\\
Department of Electrical and Computer Engineering,\\
Concordia University, Montr\'eal, QC, Canada 
\vspace*{1em}\\
\{y\_elderh,o\_hasan,tahar\}@ece.concordia.ca 
 \vspace*{3em}\\
\textbf{TECHNICAL REPORT}\\
\date{October 2019}
}
\maketitle

\newpage
\begin{abstract}
Dynamic Fault Trees (DFT) and Dynamic Reliability Block Diagrams (DRBD) are two modeling approaches that capture the dynamic failure behavior of engineering systems for their reliability analysis. Recently, two independent higher-order logic (HOL) formalizations of DFT and DRBD algebras have been developed in the HOL4 theorem prover. In this work, we propose to integrate these two modeling approaches for the efficient formal reliability analysis of complex systems by leveraging upon the advantages of each method. The soundness of this integration is provided through a formal proof of equivalence between the DFT and DRBD algebras. We show the efficiency of the proposed integrated formal reliability analysis on a drive-by-wire system as a case study.
\end{abstract}

\newpage
\section{Introduction}
\label{introduction}

Dynamic reliability models, such as dynamic  fault trees (DFTs) \cite{Merle-thesis} and dynamic reliability block diagrams (DRBDs) \cite{xu2007formal}, enable modeling the failure dependencies among system components by using DFT gates, such as Functional DEPendency (FDEP) gate, and DRBD constructs, like the spare construct. DRBDs consist of blocks that represent system components and connectors to model the successful paths or multiple paths from the input to the output. These paths determine the required system components to maintain its proper functionality. DFTs, on the other hand,  graphically model the faults of system components that lead to the failure of an undesired event, represented by a top event. The required conditions for the occurrence of this top event are captured using DFT gates.
An algebra was proposed in \cite{Merle-thesis} for the analysis of DFTs, where inputs and outputs of DFT gates are modeled based on their time of failure. In \cite{elderhalli2019probabilistic}, we developed the higher-order logic (HOL) formalization of this algebra and verified the probability of failure of commonly used DFT gates, which enables conducting a formal DFT analysis within the HOL4 theorem prover~\cite{HOL4}. \\
\indent Following the same lines of the DFT algebra, we recently proposed an algebra to analyze DRBDs with spare constructs~\cite{Yassmeen-DRBDTR}. We introduced new DRBD operators that allow expressing the structure of a given DRBD to conduct its analysis. We developed the HOL formalization of this algebra using HOL4 to ensure its soundness. It is worth mentioning that the graphical representation of the sources of failure of a system modeled as a DFT cannot be directly obtained using DRBDs. Such  a graphical representation is quite helpful in quickly identifying the vulnerabilities in systems. On the other hand, DRBDs identify the required paths and options for the successful behavior that cannot be directly identified using DFTs. However, the DRBD algebra leads to a more efficient reliability analysis since the DRBD algebra is simpler to conduct.\\ 
\indent In this work, we propose an integrated framework that enables formally analyzing DFTs and DRBDs based on their algebraic approaches. The proposed framework also allows the formal analysis of DRBDs using the DFT algebra and vice-versa, which requires verifying the formal equivalence of both algebras. The proposed integration provides the possibility to express the failure behavior of a system modeled as a DRBD and the success behavior of a system modeled as a DFT. Moreover, using the integrated framework, a given DFT can be formally modeled using the formalized DFT algebra. Then, based on the formal equivalence of the DFT and DRBD algebras, we can obtain the corresponding DRBD model of the given system in a sound manner and thus use the DRBD model to conduct the formal reliability analysis. As an illustration, we formally analyze the reliability analysis of a drive-by-wire (DBW) system \cite{altby2014design} using both reliability models and show that the DRBD algebra based formal analysis results in a shorter proof script and a smaller number of proof goals, and thus a reduction in the time required to conduct the analysis (by $1/24$ for the DBW system).

\vspace{50pt}

\section{DFT Algebra and its HOL Formalization}
\label{sec:DFT-algebra}

The algebraic approach of DFT analysis relies on presenting the  basic events, which represent system components, and the output of DFT gates based on their time of failure~\cite{Merle-thesis}. Identity elements are defined to express two states of system components. The ALWAYS element represents a component that already failed, i.e., the time of failure equals $0$. The NEVER element models a fail safe component, which means that its time of failure equals $+\infty$. Three temporal operators are also introduced, i.e., \textit{Before} ($\lhd$), \textit{Simultaneous} ($\Delta$) and \textit{Inclusive-before} ($\unlhd$), to model the dynamic behavior of one event failing before the other, at the same time and before or at the same time, respectively~\cite{Merle-thesis}. In \cite{elderhalli2019probabilistic}, we provided the HOL formalization of these operators (Table~\ref{table:element-operator}), where we defined them as lambda abstracted functions that return extended-real numbers (\texttt{extreal}), which include real numbers and $\pm\infty$ to model the NEVER element. \\

\begin{table}[!t]
 
\centering
\small
\caption{Definitions of DFT Temporal Operators}
\label{table:element-operator}
\begin{tabular}{|l|l|l|}
\hline
Operator & Mathematical Expression  &  Formalization \\ \hline  \hline
{ {\texttt{Before}}}&
$\!\begin{aligned}[b]
	{{ A  \lhd B= }{  
	\begin{cases}  A, &  A < B\\   +\infty, &  A\geq B
\end{cases}} 
}
	\end{aligned}$& $\!\begin{aligned}[c]
	&  {\texttt{$\vdash$ $\forall$ A B. 
		D\_BEFORE A B =}}\\[-1\jot]
		& {\texttt{ ($\lambda$s. if A s  < B s then A s}}\\[-1\jot] 
		&{\texttt{ else PosInf)
}}\end{aligned}$   
 \\ \hline
{ {\texttt{Simultaneous}}}& $\!\begin{aligned}[b]
	{{  A  \Delta B= }{  
	\begin{cases}   A, &  A = B\\   +\infty, &  A\neq B
\end{cases}} 
}
	\end{aligned}$ & $\!\begin{aligned}[c]
	&  {\texttt{$\vdash$ $\forall$ A B. 
		D\_SIMULT A B =}}\\[-1\jot]
		& {\texttt{ ($\lambda$s. if A s  = B s then A s}}\\[-1\jot] 
		&{\texttt{ else PosInf)
}}\end{aligned}$    \\ \hline
{ {\texttt{Inclusive Before}}}& $\!\begin{aligned}[b]
	{{  A  \unlhd B= }{	\begin{cases}   A, &  A \leq B\\   +\infty, &  A > B
\end{cases}} 
}
	\end{aligned}$ & $\!\begin{aligned}[c]
	&  {\texttt{$\vdash$ $\forall$ A B. 
		D\_INCLUSIVE\_BEFORE A B =}}\\[-1\jot]
		& {\texttt{ ($\lambda$s. if A s  $\leq$ B s then A s}}\\[-1\jot] 
		&{\texttt{ else PosInf)
}}\end{aligned}$    \\ \hline
\end{tabular}
\vspace{20pt}
\end{table}

\begin{figure}[!b]
\centering
\subfigure[OR]{
 \makebox[0.2\textwidth]{
\includegraphics[scale=0.55]{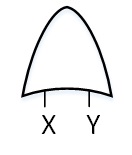}
}}
\subfigure[AND]{
 \makebox[0.18\textwidth]{
\includegraphics[scale=0.55]{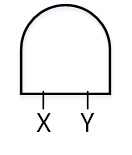}}}
\subfigure[FDEP]{
 \makebox[0.18\textwidth]{
\includegraphics[scale=0.55]{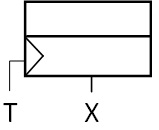}}}
\subfigure[PAND]{
 \makebox[0.18\textwidth]{
\includegraphics[scale=0.55]{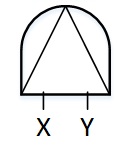}}}
\subfigure[Spare]{
 \makebox[0.18\textwidth]{
\includegraphics[scale=0.55]{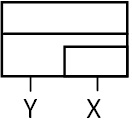}}}
\caption{Fault Tree Gates}
\label{fig:DFT_Gates}
\end{figure}

\indent In \cite{Merle-thesis}, the DFT gates, shown in Figure~\ref{fig:DFT_Gates}, are modeled based on the time of failure of their output. For instance, the Functional DEPendency (FDEP) gate is used to model failure triggers of system components. The spare gate models spare parts in a system, where the spare ($X$) replaces a main part ($Y$) after its failure. In the general case, the failure distribution of the spare is attenuated by a dormancy factor from the active state. Therefore, in the DFT algebra, two variables are used to distinguish the spare in both its states; active ($X_{a}$) and dormant ($X_{d}$). Table~\ref{table:gates} lists the definitions of these gates. In~\cite{elderhalli2019probabilistic}, we provided the HOL formalization of these gates. However, to verify the probability of failure expression given in Table~\ref{table:gates}, it is required first to define a \texttt{DFT\_event} to be used in the probabilistic analysis. This is formally defined as\cite{elderhalli2019probabilistic}:

\begin{defn}
\small{\texttt{$\vdash\forall$p X t. DFT\_event p X t = \{s | X s $\scriptstyle\leq$ Normal t\} $\cap$ p\_space p} }
\end{defn}

\noindent where \texttt{p} is a probability space. \texttt{p\_space} is a function that returns the space of \texttt{p}. \texttt{X} is the time to failure function that can represent inputs and outputs of DFT gates and \texttt{t} is the time until which we are interested in finding the probability of failure. The type of \texttt{t} is real, while the time to failure functions are of type \texttt{extreal} and thus it is required to typecast \texttt{t} to \texttt{extreal} using the \texttt{Normal} function.
We verified the probability of failure of all DFT gates based on this event and using their formal definitions, as given in Table~\ref{table:gates}~\cite{elderhalli2019probabilistic}.\\

\begin{table}[!t]
\centering
\small
\caption{DFT Gates Expressions and Probability of Failure}
\label{table:gates}
\begin{tabular}{|l|l|l|}
\hline
Gate                   & Mathematical Expression & Probability of Failure \\ \hline \hline
 {AND}                    & $\!\begin{aligned}[b]
	{{  X \cdot Y = max (X,Y)
} 
}
	\end{aligned}$ &$\!\begin{aligned}[b]
	{{  F_{X}(t) \times F_{Y}(t)
} 
}
	\end{aligned}$  \\ \hline
 {OR }                                                           & $\!\begin{aligned}[b]
	{{  X + Y = min (X,Y)
} 
}
	\end{aligned}$ & $\!\begin{aligned}[b]
	{{  F_{X}(t) + F_{Y}(t) - F_{X}(t) \times F_{Y}(t)
} 
}
	\end{aligned}$  \\ \hline
 {PAND}                  & $\!\begin{aligned}[b]
	{{  Q_{PAND}= }{ { 
	\begin{cases}   Y, &  X \leq Y\\ + \infty, &   X > Y
\end{cases}}} 
}
	\end{aligned}$  & $\!\begin{aligned}[b]
	{{  \int_{0}^{t}f_{Y}(y)\ F_{X}(y)\ dy}}
	\end{aligned}$ \\ \hline
 {FDEP}                                                 & $\!\begin{aligned}[b]
	{{  X + Y = min (X,Y)
} 
}
	\end{aligned}$   & $\!\begin{aligned}[b]
	{{  F_{X}(t) + F_{Y}(t) - F_{X}(t) \times F_{Y}(t) }}
	\end{aligned}$ \\ \hline
Spare & $\!\begin{aligned}[c]
	  Q_{SP} = & Y\cdot(X_{d} \lhd Y)+ X_{a}\cdot(Y \lhd X_{a}) \\ & {+Y\Delta X_{a}+Y\Delta X_{d}
} 

	\end{aligned}$                & $\!\begin{aligned}[c]
	&{  \int_{0}^{t} \Big{(}\int_{v}^{t} f_{(X_{a}|Y=v)} (u) du \Big{)} f_{Y}(v) dv +}\\[-2\jot]&{ { \int_{0}^{t} f_{Y}(u) F_{X_{d}}(u) du}} 
	\end{aligned}$                \\ \hline
\end{tabular}
\end{table}

\begin{figure}[!b]
\centering
\includegraphics[scale=0.55]{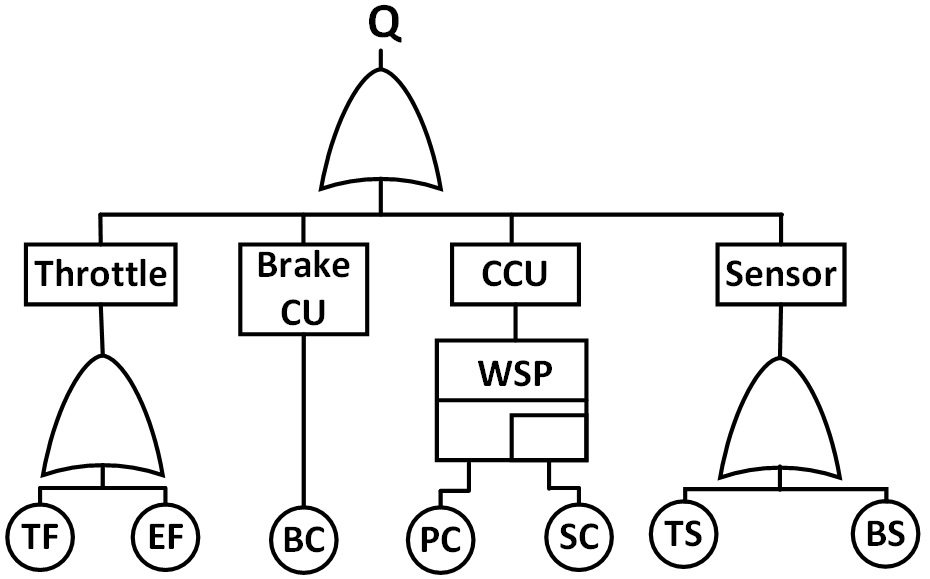}
\caption{DFT of Drive-by-wire System}
\label{fig:dbw_dft}
\end{figure}

\indent As an example, we provide the details of analyzing the DFT of a drive-by-wire system (DBW) \cite{altby2014design}, shown in Figure~\ref{fig:dbw_dft}, to explain the required steps to use our formalized algebra. This system is used in modern vehicles to control its functionality using a computerized controller. We provide the reliability model of the brake and throttle subsystems. The throttle system fails due to the failure of the throttle ($TF$) or the engine ($EF$). The brake control unit ($BCU$) failure leads to the failure of this system.  A spare gate is used to model the failure of a primary control unit ($PC$) with a warm spare ($SC$). Finally, the system can fail due to the failure of the throttle sensor ($TS$) or the brake sensor ($BS$).\\

\indent To formally conduct the analysis using our formalization, it is required first to express the function of the top event algebraically as:

\begin{minipage}{\textwidth}
\center{{\texttt{Q\textsubscript{DBW} = (TF + EF) + BCU + WSP\ PC\ SC\textsubscript{a}\ SC\textsubscript{d} + (TS + BS)}}}\\
\end{minipage}
\emph{}\\

\noindent Then, we create a \texttt{DFT\_event} for \texttt{Q\textsubscript{DBW}} as: \texttt{DFT\_event p Q\textsubscript{DBW} t}, and verify that it equals the union of the individual DFT events, i.e.:\\

\noindent {\texttt{DFT\_event p TF t $\cup$ DFT\_event p EF t $\cup$ DFT\_event p BCU t $\cup$ DFT\_event p (WSP PC SC\textsubscript{a} SC\textsubscript{d}) t $\cup$ DFT\_event p TS t $\cup$ DFT\_event p BS t}}\\

\noindent Thus, we can use the probabilistic principle of inclusion and exclusion (PIE) \cite{Merle-thesis} to verify the probability of failure of \texttt{Q\textsubscript{DBW}}. The probabilistic PIE expresses the probability of the union of events as the continuous summation and subtraction of the probabilities of  combinations of intersection of events. 
The DBW example is represented as the union of six events, therefore, applying the probabilistic PIE results in having 63 different terms in the final expression. We verify the probability of failure of the DBW as: \\

\begin{thm}
\label{thm:PROB_DBW}
\emph{}\\
\mbox{\small{\texttt{$\vdash$ $\forall$BS TS BCU PC SC$_{\texttt{a}}$ SC$_{\texttt{d}}$ EF TF p t f$_{\texttt{PC}}$ f\textsubscript{(SC\textsubscript{a}|PC)} f\textsubscript{SC\textsubscript{a}PC}. 0 $\leq$ t $\wedge$ }}}\\
\mbox{\small{\texttt{dbw\_event\_req [BS; TS; BCU; PC; SC\textsubscript{a}; SC\textsubscript{d}; EF; TF] p t f$_{\texttt{PC}}$ f\textsubscript{(SC\textsubscript{a}|PC)} f\textsubscript{SC\textsubscript{a}PC} $\Rightarrow$}}}\\
\mbox{\small{\texttt{$\bigg($prob p~(DFT\_event p Q\textsubscript{DBW} t) = }}}\\
 $\!\begin{aligned}[c]
&\small{\texttt{F\textsubscript{TF}(t)+F\textsubscript{EF}(t)+F\textsubscript{BCU}(t)+$\bigg[\int_{0}^{t}$f\textsubscript{PC}(pc)$\times\big(\int_{pc}^{t}$f\textsubscript{(SC\textsubscript{a}|PC=pc)}(sc\textsubscript{a}) $d$sc\textsubscript{a}$\big)d$pc$\bigg]$+F\textsubscript{BS}(t)+F\textsubscript{TS}}}\\[-2pt]
&\small{\texttt{-...+...- F\textsubscript{TF}(t)$\times$F\textsubscript{EF}(t)$\times$F\textsubscript{BCU}(t)$\times$F\textsubscript{BS}(t)$\times$F\textsubscript{TS}(t)$\times$}} \\[-2pt]
&\small{\texttt{$\bigg[\bigg(\int_{0}^{t}$ f\textsubscript{PC}(pc)$\times\bigg($ $\int_{pc}^{t}$f\textsubscript{(SC\textsubscript{a}|PC=pc)}(sc\textsubscript{a})$d$sc\textsubscript{a}$\bigg)d$pc$\bigg)$+$\int_{0}^{t}$f\textsubscript{PC}(pc)$\times$F\textsubscript{SC\textsubscript{d}}(pc)\ $d$pc$\bigg]\bigg)$}}
\end{aligned}$ 
\end{thm}

\noindent where \texttt{dbw\_event\_req} ensures the required conditions for independence of the events and defines the conditional density functions with their proper conditions \cite{ifm-short-code}. The first six terms in the conclusion of Theorem \ref{thm:PROB_DBW} represent the probabilities of the six individual events of the union of the DBW. Since there are 63 different terms, we are only showing a part of the theorem and the full version is available at \cite{ifm-short-code}. The script of the DBW DFT analysis required around 4850 lines of code and 24 man-hours to be developed. 

\section{DRBD Algebra and its HOL Formalization}
\label{sec:DRBD-algebra}

DRBDs capture the dynamic dependencies among system components using DRBD constructs, such as the spare and load sharing constructs. The blocks in a DRBD can be connected in series, parallel, series-parallel and parallel-series. 
Recently, we proposed an algebra that allows expressing the structure of a given DRBD based on system blocks \cite{Yassmeen-DRBDTR}. The reliability of a given system can be expressed using this DRBD algebra. We defined several operators that enable expressing DRBDs of series and parallel configurations and even more complex structures. Furthermore, the defined operators allow modeling a DRBD spare construct to capture the behavior of spares in a system. We provided the HOL formalization of this algebra to ensure its soundness and enable the formal analysis using HOL4. We first formally define a DRBD event that creates the set of time until which we are interested in finding the reliability~\cite{Yassmeen-DRBDTR}:

\begin{defn}
\small{\texttt{$\vdash\forall$p X t. DRBD\_event p X t = \{z | Normal t < X s\} $\cap$ p\_space p}}
\end{defn}

\noindent where $X$ is the time to failure function of a system component and $t$ is the moment of time until which we are interested in finding the reliability of the system. The probability of this event represents the reliability of the system until time $t$~\cite{Yassmeen-DRBDTR}:


\begin{defn}
\small{\texttt{$\vdash\forall$p X t. Rel p X t = prob p (DRBD\_event p X t)}}
\end{defn}

\noindent Then, we verify that its probability is related to the CDF~\cite{Yassmeen-DRBDTR}.

\begin{table}[t]
\caption{Definitions of DRBD Operators}
\small
\centering
\label{table:DRBD-element-operator}
\begin{tabular}{|l|l|l|}
\hline
Operator & Mathematical Expression  &  Formalization \\ \hline  \hline
{ {\texttt{AND}}}&
$\!\begin{aligned}[b]
	{{  X \cdot Y= min (X ,Y)}}	\end{aligned}$& $\!\begin{aligned}[c]
	&  {\texttt{$\vdash$ $\forall$X Y. 
		R\_AND X Y =}}\\[-1\jot]
   &	{\texttt{($\lambda$s. min (X s) (Y s))}}\end{aligned}$   
 \\ \hline
{ {\texttt{OR}}}&
$\!\begin{aligned}[b]
	{{  X + Y= max (X, Y)}
}
	\end{aligned}$& $\!\begin{aligned}[c]
	&  {\texttt{$\vdash$ $\forall$X Y. 
		R\_OR X Y =}}\\[-1\jot]
   &	{\texttt{($\lambda$s. max (X s) (Y s))
}}\end{aligned}$   
 \\ \hline

{ {\texttt{After}}}&
$\!\begin{aligned}[b]
	{{  X \rhd Y= }{  
	\begin{cases}  X, &X > Y\\ +\infty, &X\leq Y
\end{cases}} 
}
	\end{aligned}$& $\!\begin{aligned}[c]
	&  {\texttt{$\vdash$ $\forall$X Y. 
		R\_AFTER X Y =}}\\[-1\jot]
		& {\texttt{($\lambda$s. if Y s  < X s then X s}}\\[-1\jot]
   &	{\texttt{else PosInf)
}}\end{aligned}$   
 \\ \hline
{ {\texttt{Simultaneous}}}& $\!\begin{aligned}[b]
	{{  X \Delta Y= }{ 
	\begin{cases}  X, &X = Y\\ +\infty, &X\neq Y
\end{cases}} 
}
	\end{aligned}$ & $\!\begin{aligned}[c]
	&  {\texttt{$\vdash$ $\forall$X Y. 
		R\_SIMULT X Y =}}\\[-1\jot]
		& {\texttt{($\lambda$s. if X s  = Y s then X s}}\\[-1\jot]
   &	{\texttt{else PosInf)
}}\end{aligned}$    \\ \hline
{ {\texttt{Inclusive After}}}& $\!\begin{aligned}[b]
	{{  X  \unrhd Y=}{   
	\begin{cases}  X, &X \geq Y\\ +\infty, &X < Y
\end{cases}} 
}
	\end{aligned}$ & $\!\begin{aligned}[c]
	&  {\texttt{$\vdash$ $\forall$ X Y. 
		R\_INCLUSIVE\_AFTER X Y =}}\\ 		& {\texttt{($\lambda$s. if Y s  $\leq$ X s then X s}}\\[-1\jot]
   &	{\texttt{else PosInf)
}}\end{aligned}$    \\ \hline
\end{tabular}
\vspace{20pt}
\end{table}

We introduced DRBD identity elements and operators to model both the combinatorial and dynamic behaviors, as listed in Table~\ref{table:DRBD-element-operator}. The idea is similar to the DFT algebra, where the blocks are modeled based on their time of failure. We need to recall that DRBDs are concerned in modeling the successful behavior, i.e., the ``not failing" behavior, and thus we can use the time to failure functions to model the behavior of a given DRBD. We defined two identity elements for DRBD that are similar to the DFT elements, i.e., ALWAYS = $0$ and NEVER = $+\infty$. The DRBD operators are listed in Table~\ref{table:DRBD-element-operator}. The AND operator ($\cdot$) models series DRBD blocks, where it is required that all the blocks are working. The output of the AND operator fails with the first failure of any component of its inputs. On the other hand, the OR operator ($+$) models parallel structures, where at least one of the blocks should continue to work to maintain the system functionality.  To capture the dynamic behavior, we introduced three temporal operators, i.e., \textit{After}, \textit{Simultaneous} and \textit{Inclusive-after}~\cite{Yassmeen-DRBDTR}. The after operator ($\rhd$) models the sequence of events, where the system continues to work as long as one component continues to work after the failure of the other. The simultaneous operator ($\Delta$) is similar to the one of the DFT algebra, where its output fails when both inputs fail at the same time. Finally, the inclusive-after operator ($\unrhd$) combines the behavior of both after and simultaneous operators. We provided the HOL formalization of these elements and operators based on lambda abstracted functions and \texttt{extreal} numbers. The mathematical expressions and the HOL formalization are listed in Table~\ref{table:DRBD-element-operator}. The reliability expressions of these operators are available at~\cite{Yassmeen-DRBDTR}.

\begin{figure}[!t]
\centering
\includegraphics[scale = 0.8]{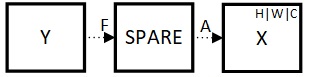}
\caption{DRBD Spare Construct}
\label{fig:drbd_spare}
\end{figure}

\indent A spare construct, shown in Figure~\ref{fig:drbd_spare}, is introduced in DRBDs to model spare parts in systems by having spare controllers that activate the spare after the failure of the main part. In Table~\ref{table:spare_reliability}, $Y$ is the main part and after its failure $X$ is activated. We use two variables ($X_{a}$, $X_{d}$), like the DFT algebra.

\begin{table}[!t]
\centering
\small
\caption{Mathematical and Reliability Expressions of Spare Constructs}
\label{table:spare_reliability}
\begin{tabular}{|c|c|}
\hline
 \small{Math. Model} & \small{Reliability}\\ \hline\hline$ {
 Q_{SP}= (X_{a} \rhd Y)\cdot (Y \rhd X_{d}) }$         &       $\!\begin{aligned}[t]  { R_{SP}(t) =} &  {1 - \int_{0}^{t} \int_{y}^{t} f_{(X_{a}|Y=y)}(x)\ f_{Y}(y) dx dy}\\& { - \int_{0}^{t} f_{Y}(y)F_{X_{d}}(y)dy } \end{aligned}$             \\ \hline
\end{tabular}
\end{table}

\begin{table}[!b]
\centering
\renewcommand{\arraystretch}{1}
\caption{Mathematical Models and Reliability of Series and Parallel Structures}
\label{table:series-parallel}
\begin{tabular}{ll|l|l|}
\cline{3-4}
                              &     & \small{Math. Model} & \small{Reliability} \\ \hline
\multicolumn{1}{|l}{$\!\begin{aligned}[c]    Series \end{aligned}$}   & {\raisebox{-0.25cm}{$\!\begin{aligned}[c]\includegraphics[scale = 0.45]{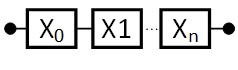}\end{aligned}$}} & $  \bigcap_{i=1}^{n}(event\ (X_{i},\ t))$  &$  \prod_{i=1}^{n}R_{X_{i}}(t)$                        \\ \hline
\multicolumn{1}{|l}{$\!\begin{aligned}[b]   Parallel \end{aligned}$} &{\raisebox{-0.9cm}{ $\!\begin{aligned}[c] \includegraphics[scale = 0.45]{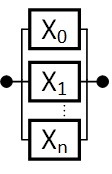}\end{aligned}$}} &    $  \bigcup_{i=1}^{n}(event\ (X_{i},\ t))$ & $  1- \prod_{1=1}^{n}(1-R_{X_{i}}(t))$    \\ \hline
\end{tabular}
\end{table}

DRBD blocks can be connected in series, parallel and more nested structures. We provide here the details of only the series and parallel structures, as listed in Table~\ref{table:series-parallel}. Details about the nested structures can be found in \cite{Yassmeen-DRBDTR}. The series structure, shown in Table~\ref{table:series-parallel}, continues to work as long as all the blocks are working. Once one of these blocks stops working, then the entire system stops as well. It can be expressed using the AND operator. Its mathematical model is expressed as the intersection of the individual DRBD events \cite{hasan2015reliability}. The parallel structure, shown in Table~\ref{table:series-parallel}, is composed of several blocks that are connected in parallel. Its structure function can be expressed using the OR operator. Its mathematical model is represented using the union of the individual DRBD events. We developed the HOL formalization of these structures and verified their reliability expressions assuming the independence of the individual blocks~\cite{Yassmeen-DRBDTR}.

\begin{figure}[!b]
\centering
\includegraphics[scale=0.8]{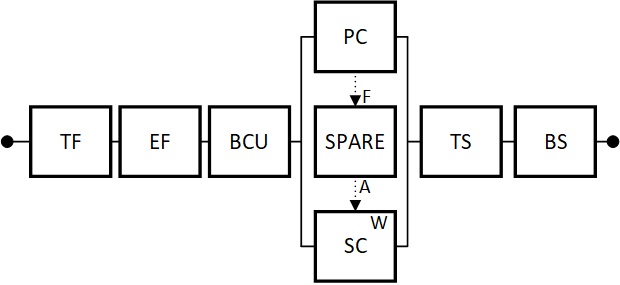}
\caption{DRBD of Drive-by-wire System}
\label{fig:dbw_drbd}
\end{figure}

We demonstrate the applicability of the DRBD algebra in the formal analysis of the DRBD of the DBW system given in Figure~\ref{fig:dbw_drbd}. This DRBD is a series sructure with one spare construct to model the main part $PC$ that is replaced by $SC$ after failure. The structure function of the DBW DRBD ($F_{DBW}$) can be expressed as:
\begin{equation}
F_{DBW} = TF \cdot EF \cdot BCU \cdot (SC_{a} \rhd PC) \cdot (PC \rhd SC_{d}) \cdot TS \cdot BS
\end{equation}
Then, we verify the reliability of the DBW system as:\\

\begin{thm}
\label{thm:Rel_DBW}
\textup{\small{\texttt{$\vdash\forall$p TF EF BCU PC SC\textsubscript{a} SC\textsubscript{d} TS BS t.}}}\\
\mbox{\textup{\small{\texttt{~DBW\_set\_req p TF EF BCU PC SC\textsubscript{a} SC\textsubscript{d} TS BS t $\Rightarrow$}}}}\\
\mbox{\textup{\small{\texttt{~(prob p (DRBD\_event p F\textsubscript{DBW} t) =}}}}\\
\mbox{\textup{\small{\texttt{~~Rel p TF t * Rel p EF t * Rel p BCU t * Rel p (R\_WSP PC SC\textsubscript{a} SC\textsubscript{d}) t *}}}}\\ \mbox{\textup{\small{\texttt{~~Rel p TS t * Rel p BS t})}}}
\end{thm}

\noindent where \texttt{DBW\_set\_req} ascertains the required conditions for the independence of the DBW system blocks~\cite{Yassmeen-DRBDTR}. The reliability of the spare construct can be further rewritten using the reliability expression of the spare using integrals. The script of the reliability analysis of the DBW DRBD is 150 lines long and required only one hour of work. \\

\section{Integrated Framework for Formal DFT-DRBD Analysis}

\begin{figure}[!b]
\centering
\includegraphics[width= \textwidth]{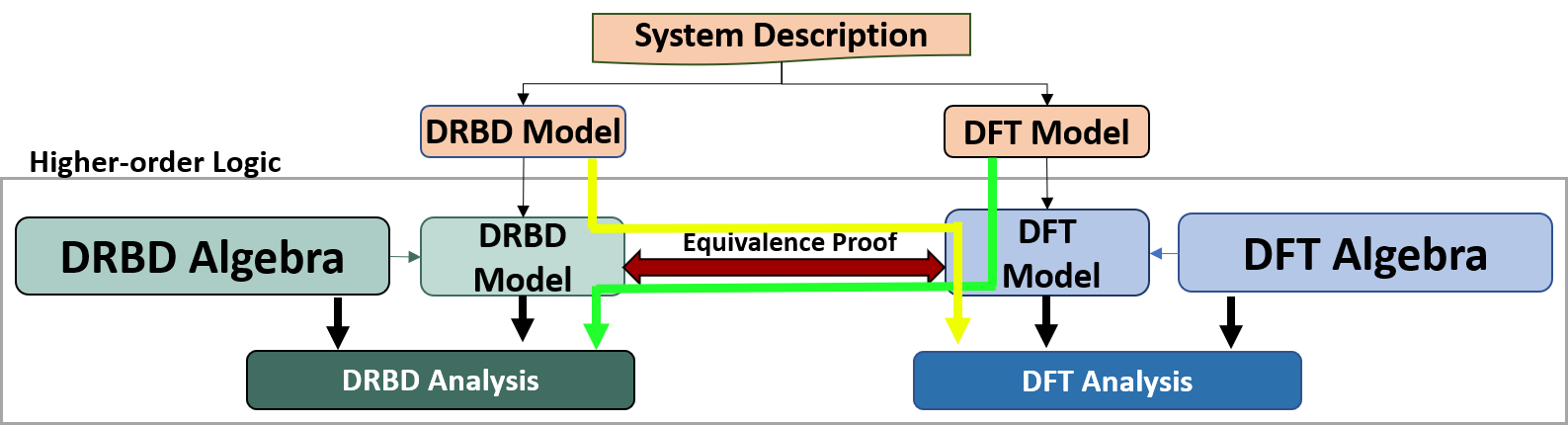}
\caption{Integrated Framework for Formal DFT-DRBD Analysis using HOL4}
\label{fig:methodology}
\end{figure}

The proposed framework integrating DFT and DRBD algebras is depicted in Figure~\ref{fig:methodology}.
It can be utilized to conduct both DFT and DRBD analyses using the HOL formalized algebras and allows formally converting a DFT model into its corresponding DRBD based on the equivalence of both algebras. The analysis starts by a given system description that can be modeled as a DFT or DRBD. Formal models of the given system can be created based on the HOL formalized algebras. The DRBD model can be analyzed as described in Section~\ref{sec:DRBD-algebra}, where a DRBD event is created and its reliability is verified based on the available verified theorems of DRBD algebra. On the other hand, a DFT model can be analyzed using the formalized DFT algebra, which requires dealing with the probabilistic PIE. Furthermore, the DRBD model can be converted to a DFT to model the failure instead of the success, then this model is analyzed using the DFT algebra. Similarly, the DFT model can be analyzed by converting it to its counterpart DRBD model, which results in  an easier process as the PIE is not invoked. 

In order to handle the DFT analysis using DRBD algebra and the DRBD analysis using the DFT algebra, it is required to be able to represent the DRBD of the corresponding DFT gates using the DRBD algebra and vice-versa (the equivalence proof in Figure~\ref{fig:methodology}). According to \cite{distefano2007dynamic}, the OR, AND and FDEP gates can be represented using series, parallel and series RBDs, respectively. Therefore, they can be modeled using AND and OR operators, while the spare gate corresponds to the spare construct. Finally, the PAND gate can be expressed using the inclusive after operator ($Y \unrhd X$). However, we need to formally verify this equivalence to ensure its correctness. In Table~\ref{table:verify-DRBD-DFT}, we provide the theorems of equivalence of DFT gates and DRBD operators and constructs, where \texttt{D\_AND}, \texttt{D\_OR}, \texttt{FDEP}, \texttt{P\_AND} and \texttt{WSP} are the names of the AND, OR, FDEP, PAND and spare DFT gates in our HOL formalization~\cite{elderhalli2019probabilistic}. \texttt{R\_WSP} is the name of the spare DRBD construct in our formalized DRBD \cite{Yassmeen-DRBDTR} and \texttt{ALL\_DISTINCT [Y X\textsubscript{a} X\textsubscript{d}]} ensures that the inputs cannot fail at the same time.

\begin{table}[!t]
\centering
\small
\caption{Verified Equivalence of DFT Gates and DRBD Algebra}
\label{table:verify-DRBD-DFT}
\begin{tabular}{|c|c|l|}
\hline
\small{DFT Gate}& \small{DRBD Operator/Construct}        & \small{Verified Theorem} \\ \hline \hline
\scriptsize{AND} &  {OR} & {\texttt{$\vdash\forall$X Y. D\_AND X Y = R\_OR X Y}}                              \\ \hline
 {OR} &  {AND} &  {\texttt{$\vdash\forall$X Y. D\_OR X Y = R\_AND X Y}}                             \\ \hline
 {FDEP} &  {AND} &  {\texttt{$\vdash\forall$X Y. FDEP X Y = R\_AND X Y}}                             \\ \hline
 {PAND}  &  {Inclusive After} & $\!\begin{aligned}[c] &{\texttt{$\vdash\forall$X Y. P\_AND X Y =}}\\[-1\jot]
 &\texttt{{ R\_INCLUSIVE\_AFTER Y X}}  \end{aligned}$                             \\ \hline
 {Spare} &  {Spare} & $\!\begin{aligned}[c]
	&  {\texttt{$\vdash\forall$X\textsubscript{a} X\textsubscript{d} Y.}}\\[-1\jot]
	&\texttt{{($\forall$s. ALL\_DISTINCT [Y s;X\textsubscript{a} s;X\textsubscript{d} s]) $\Rightarrow$ }}\\[-2\jot]
& {\texttt{(WSP Y X\textsubscript{a} X\textsubscript{d} = R\_WSP Y X\textsubscript{a} X\textsubscript{d})}}                           \end{aligned}$   
  \\ \hline
\end{tabular}
\end{table}

In order to use these verified expressions in Table~\ref{table:verify-DRBD-DFT}, we need to verify that the \texttt{DRBD\_event} and the \texttt{DFT\_event} possess complementary sets in the probability space. We formally verify this as:

\begin{thm}
\small{\texttt{$\vdash\forall$p X t. prob\_space p $\wedge$ (DFT\_event p X t) $\in$ events p $\Rightarrow$}}\\
\mbox{\small{\texttt{(prob p (DRBD\_event p X t) = 1 - prob p (DFT\_event p X t))}}}
\end{thm} 

\noindent where the conditions ensure that \texttt{p} is a probability space and that the DFT event belongs to the events of the probability space. This theorem can be verified also if we ensure that the DRBD event belongs to the probability space. This theorem means that for the same time to failure function, the DRBD and DFT events are the complements of each other. This way, we can analyze DFTs using the DRBD algebra and vice-versa. 

Based on the verification results obtained in Table~\ref{table:verify-DRBD-DFT}, DFT gates can be formally represented using DRBDs. We show that the amount of effort required by the verification engineer to formally analyze DFTs by analyzing its counterpart DRBD is less than that of analyzing the original DFT model.  In Section~\ref{sec:DFT-algebra}, a DFT is formally analyzed using the DFT algebra by expressing the DFT event of the structure function as the union of the individual DFT events. Then the probabilistic PIE is utilized to formally verify the probability of failure of the top event. The number of terms in the final result equals $2^n-1$, where $n$ is the number of individual events in the union of the structure function. Therefore, in the verification process, it is required to verify at least $2^n-1$ expressions. On the other hand, verifying a DRBD would require verifying a single expression for each nested structure.

As an example, consider the reliability analysis of the DBW system. Analyzing the DFT of this system required verifying 63 subgoals as the top event is composed of the union of six different events. While analyzing the DRBD of the DBW system required verifying only one main subgoal to be manipulated to reach the final goal. Table~\ref{table:compare} provides a comparison of the size of the script, the required time to develop it and the number of goals to be verified. Based on these observations, analyzing the reliability of the DBW using the DRBD required $1/24$ of the time needed by the DFT. These results show that it is more convenient to analyze the DRBD of a system rather than its DFT if the algebraic approaches are to be used. The only added step will be to formally verify that the DFT and DRBD are the complements of each other, which is straightforward utilizing the theorems in Table~\ref{table:verify-DRBD-DFT}. Therefore, we verify this as:\\

\begin{thm}
\small{\texttt{$\vdash\forall$p TF EF BCU PC SC\textsubscript{a} SC\textsubscript{d} TS BS t. }}\\
\mbox{\small{\texttt{prob\_space p $\wedge$ DBW\_events\_p p TF EF BCU PC SC\textsubscript{a} SC\textsubscript{d} TS BS t $\Rightarrow$.}}}\\
\mbox{\small{\texttt{(prob p (DRBD\_event p F\textsubscript{DBW} t) = 1- prob p (DFT\_event p Q\textsubscript{DBW} t))}}}\\
\end{thm}

\noindent where \texttt{DBW\_events\_p} ensures that the DBW DFT events are in the events of the probability space. Thus, we can use the DRBD reliability expression (Theorem~\ref{thm:Rel_DBW}) to verify the probability of failure of the DFT, which results in a reduction in the analysis efforts. 

\begin{table}[!t]
\centering
\caption{Comparison of Formal Analysis Efforts of DBW}
\label{table:compare}
\begin{tabular}{c|c|c|c|}
\cline{2-4}
                           & \small{\# of subgoals} & \small{\# of lines in the script} & \small{required time} \\ \hline \hline
\multicolumn{1}{|c|}{\small{DFT}}  & \small{ 63}        &   \small{ 4850}   & \small{ 24 hours}    \\ \hline
\multicolumn{1}{|c|}{\small{DRBD}} &   \small{ 1 }     & \small{150}    & \small{1 hour }     \\ \hline
\end{tabular}
\end{table}

\section{Conclusions}
In this report, we proposed an integrated framework to enable the multiway formal algebraic analysis of DFTs and DRBDs within a theorem prover. This framework allows transforming a DFT and DRBD models into their corresponding DBRD and DFT models, respectively, to be either analyzed more effectively using the DRBD algebra or to clearly observe the failure dependencies in the form of a DFT. This requires formally verifying the equivalence of both DFT and DRBD algebras.  To illustrate the efficiency and usefulness of the proposed framework, we provided a comparison of the efforts required to analyze a drive-by-wire system and the results showed that using the DRBD in the analysis instead of DFTs required verifying less goals (1:63), smaller script size (150:4850) and less time (1h:24h).
\bibliographystyle{unsrt}

\bibliography{mabiblio}
\end{document}